\documentclass{rmaa}
\usepackage{amsmath}
\usepackage{xcolor}

\newcommand{\eb}{\begin{equation}}
\newcommand{\ee}{\end{equation}}

\newcommand{\masyr}{mas yr$^{-1}$}
\definecolor{rkka}{RGB}{219,66,32}

\begin{document}
\title{The Hipparcos Pleiades parallax error is also a proper motion error}

\author{Valeri V. Makarov
\affil{U.S. Naval Observatory, 3450 Massachusetts Ave., Washington, DC 20392-5420, USA}
}
\email{valeri.makarov@gmail.com, valeri.v.makarov.civ@us.navy.mil}

\resumen{}
\abstract{
The mean parallax of the Pleiades open cluster from the Hipparcos catalog, which is larger than the true value
by approximately 1 mas, was one of the greatest controversies in the area of space astrometry. The origin of
this error, as well as a possible algorithm of correcting it, was proposed by Makarov (2002). The problem
emerged from the adopted least-squares fitting algorithm of the along-scan attitude parameter, reconstructed from
the satellite's own astrometric measurements in the two simultaneously registered fields of view. The Pleiades
stars observed in one of the fields obtained a much greater weight in the attitude solution than the target stars
in the other field, owing to their brightness and number density. This created a weak condition of the least-squares fit
and enhanced the propagation of position-correlated attitude perturbations through solution iterations. An ad hoc remedy
solution using the published Intermediate Astrometry Data resulted in corrections to the mean parallax, proper motion in right ascension and declination. The accurately determined differences 
Gaia EDR3 $-$ Hipparcos for 52 common stars are close to these corrections within the formal uncertainties for all three parameters, which strongly suggests that
the proposed interpretation was correct. When corrected for the systematic vector field fitted with 126 vector spherical harmonics to degree 7, these differences amount to $(+0.39,\,-0.74)$ mas yr$^{-1}$. The implications of small-scale proper motion and position errors in Hipparcos for present day astrometry are briefly discussed.
}

\keywords{
astrometry --- reference systems --- proper motions --- parallaxes}
\shorttitle{The Hipparcos Pleiades parallax and proper motion}
\shortauthor{Valeri V. Makarov}
\maketitle

\section{Introduction}
\label{sec:intro}

The publication of the Hipparcos star catalog and its supporting data sets \citep{esa} marked the beginning
of a new golden age of astrometry and fostered even bolder ideas setting more ambitious goals. The impact of this
space mission across the diverse areas of astronomy and astrophysics is summarized by \citet{per}. The
advent of space astrometry was not incontrovertible, however. Perhaps, the most contentious issue discussed
for at least a decade was the conspicuous discrepancy between the mean trigonometric parallax of the Pleiades
open cluster and the previously adopted distance value, which had crystallized from numerous observational and
theoretical investigations \citep{mer}. The Hipparcos parallax is too large at approximately 8.6 mas, placing
the Pleiades main sequence about 0.5 mag fainter with respect to the well-established main sequence at its
metallicity. After careful analysis of possible astrophysical causes such as incorrectly determined reddening,
abnormal helium abundance, a systematic error of $\sim 1$ mas in the Hipparcos data was proposed, which is 
well outside of the $3\sigma$ statistical range \citep{pin, sod}. The first important clue toward a consistent explanation
of the discrepancy was provided by \citet{nar}, who used the convergent point method and the resulting kinematic
parallaxes (significantly more precise than the trigonometric parallaxes) for the Pleiades and Hyades to reveal
the presence of {\it sky-correlated} errors in the Hipparcos data on angular scales $2^\circ$--$3^\circ$. 

The presence of such a large error in Hipparcos astrometry had been initially doubted by some authors
privy to the mission \citep{vL99,1999A&A...345..471R}. Very soon, mounting evidence to the contrary of 
astrometric \citep{2000ApJ...533..938G,2004Natur.427..326P,2004A&A...425L..45Z,2005AJ....129.1616S,2014Sci...345.1029M,2017A&A...598A..48G} and astrophysical nature \citep{2001A&A...374..105S,2004A&A...418L..31M,2005A&A...429..887P,2006MmSAI..77..455F,2007ApJ...655..233A,2013PASP..125.1297M,2016ApJS..222...19K,2016A&A...595A..59M} was provided. One could say that the bottom line of this discussion was drawn by the Gaia astrometric mission \citep{2016A&A...595A...1G} of much superior precision, which already in its first data release \citep[DR1,][]{2016A&A...595A...2G} unambiguously confirmed that both the principal Hipparcos catalog and its later re-reduction
failed to provide an accurate mean parallax of the Pleiades. A very accurate trigonometric parallax of the Pleiades was also obtained from Gaia Data Release 2 \citep{2018A&A...616A..10G}.

One of the goals of this paper is to show that the correct interpretation of the origin of this spectacular
error has been presented by \citet{mak02}. The topic is revisited because Hipparcos continues to be
important for many studies related to positions and motions of brighter stars, as well as for practical
applications \citep{2021FrASS...8....9K}. Understanding past mistakes is also essential for planning of
future space astrometry missions, and generally, of large astrometric catalogs. The other objective is to draw the attention of the research community to the fact that Hipparcos proper motions of the brighter stars contain significant position correlated errors at a range of angular scales. These proper motions are being used in a number of important projects, for example, for detecting accelerating astrometric binaries with unresolved or dim companions \citep[e.g.,][]{ker}.

\section{Locally weak condition}
\label{con.sec}
The main principles of self-calibrating all-sky space astrometry, originally formulated by P. Lacro\^ute
in 1966 \citep[see ][]{1984MitAG..62...63K} are based on the concept of a telescope with split viewing
directions separated by a fixed ``basic angle" $\Gamma$. One-dimensional positions of stars (called abscissae) are measured on the detector as they drift across the two fields of view. Linearized condition equations include
only small differences ``observed minus calculated" of the measured abscissae and the derived astrometric parameters are therefore corrections to a set of nominal values. Likewise, the results for other unknowns
involved in this adjustment are also small corrections to some previously estimated or assumed functions of time.
Without the basic angle, the condition equations would be nearly degenerate to a wide range of perturbations.
The ability to reference each target star to a large number of other stars observed almost simultaneously at a
large angular distance is pivotal in improving the condition of space astrometry equations. Trigonometric parallaxes,
in particular,
become absolute \citep{1998A&A...340..309M}, while the propagation of large-scale sky-correlated errors of proper
motions and positions (outside of the well-known 6--rank deficiency) is greatly reduced \citep{2012AJ....144...22M}.

This theoretical advantage had yet to be realized in a carefully designed data processing pipeline where the tasks of instrument calibration, attitude reconstruction, and astrometric adjustment were divided into stages of a complex iterative process and performed by different teams. The one-dimensional along-scan attitude was modeled by piecewise cubic spline functions of time. The characteristic scale of these functions was a few degrees, matching the angular diameter of a nearby cluster. As explained in detail in \citep{mak02}, setting the weights of individual measurements depending on the measured flux gives rise to a strong imbalance between the data from the two fields of view and a locally weak condition. An imprecise initial assumption about the true abscissae of the Pleiades, for example, would not be corrected because of a poor convergence of the iterative adjustment. The proposed fix was to re-estimate the instantaneous position of the bisector assigning a greater weight to the fainter stars simultaneously observed in the other field of view.

\section{The mean Pleiades parallax and proper motion in Gaia EDR3}
\label{473.sec}

Gaia EDR3 \citep{2021A&A...649A...1G} includes many more members of the Pleiades cluster than Hipparcos because of the fainter magnitude limit. For the purpose of unbiased comparison, however, we have to use the same sample of stars as in \citet{mak02}. It includes 53 bona fide Pleiades members with $H_p$ magnitudes between 2.85 and 10.87 and a median magnitude 8.32. The brightest star, Alcione = HIP 17702, however, does not have proper motion or parallax determinations in Gaia EDR3. It is eliminated from the analysis leaving 52 stars. Some of these stars have fainter companions within $10\arcsec$ resolved in EDR3, with three companions being likely members of the cluster and possibly binary companions according to the astrometric information. In each case, the closer match is much brighter removing any ambiguity in cross-identification. 

The mean differences of parallax and proper motion components, $\Delta \varpi$, $\Delta \mu_{\alpha *}$, and $\Delta \mu_{\delta}$, are computed in the sense ``Gaia $-$ Hipparcos" with weights $w_i$ inversely proportional to the combined formal variances as given in the two catalogs, e.g.,
\eb
\Delta \varpi = \frac{\sum_{i=1}^{52}(\varpi_{{\rm EDR3},i}-\varpi_{{\rm HIP},i})/(\sigma^2_{\varpi_{{\rm EDR3},i}}+
\sigma^2_{\varpi_{{\rm HIP},i}})}{\sum_{i=1}^{52}(\sigma^2_{\varpi_{{\rm EDR3},i}}+
\sigma^2_{\varpi_{{\rm HIP},i}})^{-1}}.
\ee
The standard error of the mean is then computed as
\eb
\sigma_{\Delta \varpi} = \left({\sum_{i=1}^{52}(\sigma^2_{\varpi_{{\rm EDR3},i}}+
\sigma^2_{\varpi_{{\rm HIP},i}})^{-1}}\right)^{-\frac{1}{2}},
\ee
and the same formulae are applied to the proper motion components $\mu_{\alpha*}$ and $\mu_\delta$.

In application to proper motions, this is a simplified way of estimating statistical differences, which includes only the coordinate projections of the proper motion vector $\mu_{\alpha *}$ and $\mu_\delta$.
A mathematically more consistent method is to compute for each star the vector differences ${\boldsymbol \delta}_\mu=\boldsymbol{\mu}_{\rm EDR3}-\boldsymbol{\mu}_{\rm HIP}$ and their normalized bivariate values
\eb
\chi_\mu=\left({\boldsymbol \delta}_\mu^T\; {\boldsymbol C}_\mu^{-1}\; {\boldsymbol \delta}_\mu\right)^\frac{1}{2},
\ee
where ${\boldsymbol C}_\mu$ is the sum of the corresponding $2\times 2$ covariance matrices of the proper motion vectors in Gaia and Hipparcos. The statistics $\chi_\mu$ are expected to be distributed as $\chi(1)$. For an adequate comparison with the results from \citep{mak02}, however, the simplified method is required because the full covariance matrix of the updated Hipparcos proper motion was not available.

\begin{figure}\centering
\includegraphics{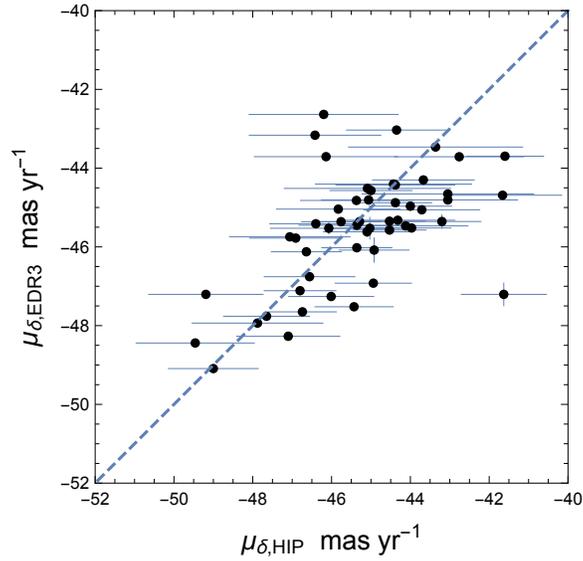}
\caption{Declination components of proper motions of 52 Pleiades members in Gaia EDR3 and Hipparcos. 
\label{dmu.fig} }
\end{figure}

We note that the weights of proper motion component differences are dominated by the uncertainties in Hipparcos, where the formal errors are much greater than in Gaia EDR3. This is seen from the error bars in Fig. \ref{dmu.fig}, which shows the $\delta$ components of the 52 proper motions in EDR3 plotted versus the corresponding values in Hipparcos. The straight diagonal line shows the location of data points when the determinations in the two catalogs are ideally consistent. More points are shifted below this line, reflecting the significant difference between the mean proper motions.

The mean differences thus computed are $\Delta \varpi=-1.00\pm 0.17$ mas, $\Delta \mu_{\alpha *}=+0.34\pm 0.17$ \masyr, and $\Delta \mu_{\delta}=-0.62\pm 0.13$ \masyr. The corresponding confidence bounds for these mean values assuming a normal distribution of errors are 1, 0.9772, and 1, respectively. These values are statistically close to the corrections
intrinsically derived by \citet{mak02} from the Hipparcos data: $-0.71\pm 0.14$ mas, $+0.45\pm 0.14$ \masyr, 
$-0.66\pm 0.11$ \masyr, respectively. The formal confidence bounds for these estimates are 1, 0.9993, and 1, respectively. We note that the probability of the null hypothesis that these updates coincide within $1.7\cdot\sigma$ or better in all the three parameters by accident is practically zero.

The Gaia EDR3 and Hipparcos proper motion systems for brighter stars are known to have systematic, or large-scale sky-correlated differences. They include a substantial rigid 3D rotation, which can be viewed
as a spin of the entire frame with an angular acceleration. The pattern of Gaia$-$Hipparcos proper motion differences, however, is not limited to this rigid rotation. To estimate the contribution of the large-scale vector field to the Pleiades-specific proper motion differences, a dedicated analysis of the proper motion systems
was performed. The Hipparcos catalog positions and their covariances were transferred onto the mean epoch of Gaia DR3 (2016) and all common stars were crossmatched. After removal of a large number of known binary and double stars, as well as statistically perturbed unresolved astrometric binaries, some $75,000$ well-behaved stars with proper motions in both catalogs remained. A set of 126 vector spherical harmonics (which is a complete set up to degree 7) was fitted to the observed Gaia$-$Hipparcos vector field using the formal covariance matrices for optimal weights. Each vector spherical harmonic represents a specific vector filed pattern on the sky, and its coefficient defines the estimated amplitude. The result revealed a spectrum of harmonics with statistically significant signal-to-noise ratios, where the three magnetic (or, toroidal) harmonics of degree 1 were by far the greatest contributors. The median vector length of the fitted field is 191 $\mu$as yr$^{-1}$. At the location of the Pleiades cluster, the fitted value is $(-48,\, +123)$ $\mu$as yr$^{-1}$. In order to correct for the large-scale distortion pattern, this vector should be subtracted from the above quoted mean proper motion differences. The result $(+0.39,\, -0.74)$ mas yr$^{-1}$ is marginally closer to the estimates from \citet{mak02}.

\section{Discussion}
\label{con.sec}

For nearly two decades, the Hipparcos catalog provided an optical realization of the International Celestial Reference System. Its emergence motivated fundamental changes in the definition of the celestial reference frame with a decisive move from the dynamic mean equinox of J2000 to a mathematically fixed, precession-free origin on the similarly defined equator \citep{2002A&A...392..341S}. The endeavor of achieving a maximally inertial (non-rotating) reference system of celestial coordinates resulted in the shift of paradigm from referencing stellar positions to the solar system ephemerides. The valued quasi-inertiality is based on a complex and indirect link to the extragalactic reference frame \citep{1997A&A...323..620K}. This link is much more robust and straightforward for Gaia, which observed $\sim 10^6$ optical quasars and AGNs, as well as a few thousand optical counterparts of ICRF3 \citep{2018A&A...616A..14G}. Unfortunately, because of essentially separate calibration pipelines for bright stars and fainter sources, this link does not fully apply to the sample of Hipparcos stars. This may be one of the reasons for significant global and large-scale proper motion differences. Gaia EDR3 proper motions
of brighter stars, in particular, may have significant rigid spin components with respect to the better
constrained fainter stars \citep{2021A&A...649A.124C}.

Hipparcos astrometry acquires a pivotal role of the first-epoch realization of a quasi-inertial celestial reference frame. Any distortions of its position and proper motion systems affect the long-term viability of this frame. One application is the detection of $\Delta\mu$ binaries \citep{wie}, which are unresolved astrometric binaries with variable proper motion. The increased sensitivity of this method allows us to detect Jupiter-mass planets in
long-period orbits around nearby stars. It is also relevant for maintaining the optical reference frame, because only truly single stars can be stable astrometric standards. 
The present follow-up study shows that Hipparcos parallaxes and proper motions are burdened by small-scale error of technical origin, which is, in principle, correctible. It is likely that the mean positions in Hipparcos are also affected in certain parts of the celestial sphere by similar or larger amounts. Such errors are bound to propagate into the most precise Hipparcos$-$Gaia (HG) proper motions, which are central to the anticipated detection of long-period companions of planetary mass or inactive black hole companions of stellar mass \citep{2017MNRAS.470.2611M}. This error overhead coming from unaccounted perturbations of Hipparcos positions may exceed the intrinsic formal error of short-term Gaia proper motions, so its presence is bound to perturb the detection results toward increased occurrence of false positives at the most affected locations. Ongoing investigations of the Gaia$-$Hipparcos proper motion field should at a minimum take into account possible sky-correlated deviations. A more ambitious goal would be to devise ways of correcting Hipparcos astrometry once we have understood the origin of its main weakness. A complete reprocessing of Hipparcos mission data starting with the Intermediate Astrometry Data (HIAD) and using Gaia results for certified stable and single astrometric standards can be considered. Such a reprocessing effort should focus on improving the attitude, basic angle $\Gamma$, and reference great circle zero-points solutions. The technical feasibility of such a solution for the latter data type has been demonstrated by \citet{2022arXiv220409080Z} using precision astrometric data from modern ground-based observations. In the same paper, the technical possibility of reprocessing the available Hipparcos data and solving for an improved set of abscissae zero-points has been demonstrated. Gaia positions of the fainter Hipparcos stars can be extrapolated back to 1991.25 using corrected Gaia proper motions and used as additional constraints in such a global solution, potentially improving the system of Hipparcos positions on the largest spatial scale including rotation.
Another possibility is to improve the Hipparcos proper motion system {\it a posteriori} at the cost of removing the option of external verification of Gaia. The main obstacle on this way is the existence of similar sky-correlated errors in Gaia, including a possible rigid spin of the entire proper motion system already seen in the Gaia Data Release 2 \citep{2019gaia.confE..25M}. Although these imperfections are much smaller in Gaia than in Hipparcos, they create ambiguity in the interpretation of obvious differences between the two catalogs. A deeper insight into the propagation of correlated errors in Hipparcos may help to disentangle this ambiguity.

\section*{Acknowledgments}

The anonymous referee is thanked for useful and constructive suggestions.
This work has made use of data from the European Space Agency (ESA) mission
{\it Gaia} (\url{https://www.cosmos.esa.int/gaia}), processed by the {\it Gaia}
Data Processing and Analysis Consortium (DPAC,
\url{https://www.cosmos.esa.int/web/gaia/dpac/consortium}). Funding for the DPAC
has been provided by national institutions, in particular the institutions
participating in the {\it Gaia} Multilateral Agreement.

\end{document}